# Webpage Segmentation for Extracting Images and Their Surrounding Contextual Information

F. Fauzi, H. J. Long, M. Belkhatir
School of IT, Monash University

## ABSTRACT
Web images come in hand with valuable contextual information. Although this information has long been mined for various uses such as image annotation, clustering of images, inference of image semantic content, etc., insufficient attention has been given to address issues in mining this contextual information. In this paper, we propose a webpage segmentation algorithm targeting the extraction of web images and their contextual information based on their characteristics as they appear on webpages. We conducted a user study to obtain a human-labeled dataset to validate the effectiveness of our method and experiments demonstrated that our method can achieve better results compared to an existing segmentation algorithm.

## Categories and Subject Descriptors
H.3.1 [**Information Storage and Retrieval**]: Content Analysis and Indexing – *indexing methods*

## 1. INTRODUCTION
As the World Wide Web fuses into our existence, an abundance of images can be found on the Web. Incidentally, these Web images come with rich contextual information, which is the text associated to the images, used jointly with their filename, alt description, and page title.

This contextual information has varying definitions and has been perceived as a window of words [14], a paragraph [5, 15], a section [4, 9, 12] and even the entire page [6, 7].

There are two general methods for extraction of image contextual information. The first and simplest method is to use a fixed window size (min: 20 terms to max: entire page) whereby a fixed number of words before and after the image are considered as the image surrounding context. Alternatively, the second method performs webpage segmentation to extract sections containing the images and their surrounding context [1, 4, 9, 12]. Webpage segmentation is the task of breaking a webpage into sections that appear coherent to a user browsing the webpage as will be further discussed in the next section.

Both are not without problems. The first method, although straightforward, tends to produce low-level accuracy as texts tend to be associated to the wrong image, for instance, when the image description appears only after the image. And when taking the entire page, the surrounding context will contain too much noisy information.

As for the second method, we believe that webpage segmentation is the natural method for extracting image surrounding context. Nevertheless, there are problems that need addressing: i) the ambiguity in defining the boundary of the contextual information of each image ii) the heterogeneity of webpages – different websites having different content layout iii) the parameters/modifications required to tune general webpage segmentation algorithm to extract images and their surrounding context and iv) the performance of the segmentation algorithm in terms of time required to process a webpage, extract images and their corresponding surrounding context, a fast algorithm would be required to cater to the large and growing number of images of the Web.

**Our Contributions**. To address these concerns, we propose a fast DOM Tree-based segmentation algorithm that does not require any tuning parameters, targeting the extraction of images and their surrounding context, which we refer as image segments and test it against a human-labeled dataset obtained via a user study. Our method can extract image segments from a diverse range of websites, thus making it practical and scalable. Experimental results indicate that our method outperforms an existing state of the art segmentation algorithm, VIPS [2] in precision and recall.

## 2. RELATED WORK
Efforts to segment webpages for extracting surrounding context can be categorized into two: i) DOM Tree-based and ii) DOM Tree-based with additional visual information obtained from rendering the DOM Tree.

Typically, the webpage DOM tree structure is analyzed to discover segment-specific patterns. [5, 15] extract a paragraph of texts containing the image. Hua *et al.* [9] rely on the border properties of structural HTML markup elements such as <TABLE>, <TR>, <TD>, <DIV> and <HR>. Feng *et al.* [4] consider these structural tags as separators and have a cutoff point at text description length greater than 32 words before and after an image. While efficient, the heuristics used above work on limited webpages, and [4] fell back on fixed window size. Hence, better heuristics should be used to improve scalability to various types of webpages.

Cai *et al.*'s Vision-based Page Segmentation (VIPS) algorithm [2] is a general webpage segmentation algorithm that uses visual information obtained from rendering the webpage, in addition to the DOM tree structure. [1, 13] implement VIPS for the extraction of image surrounding context by reducing webpages to image blocks and taking all texts within a block as the surrounding context. The major problem in VIPS is the value of the Permitted Degree of Coherence (PDoC), which ranges from 1-10 and defines the different granularities of the segmentation algorithm to cater for different applications. In [8], the PDoC is empirically set to 5, while this may work for some pages; generally it takes more contextual information than required by considering a bigger section encompassing an image. Increasing the PDoC would cause an opposite effect. Li *et al.* [12] too include visual cues of size and position in their page segmentation algorithm. Even though visual cues might improve accuracy, these algorithms are known to be computationally expensive and become crucial when processing the large-scale Web.

Other webpage segmentation algorithms that have been developed to address information retrieval applications are reviewed. Kao *et al.*[10] separate blocks of the DOM sub-trees by comparing the entropies of the terms within the blocks. Chakrabarti *et al.*'s meta-heuristic Graph-Theoretic approach [3] cast the DOM tree as a weighted graph, where the weights indicate if two DOM nodes should be placed together or in different segments and Kohlschutter *et al.* [11] applied quantitative linguistics and computer vision strategies to the segmentation problem. These segmentation algorithms would require further modifications to suit our purpose.

## 3. FORMULATION
### 3.1 Characteristics of Web Images
Our observation on Web images embedded within webpages sampled from business, shopping, governmental, education, news and informational sites shows three classes of Web images irrespective of webpage category – unlisted, listed and semi-listed images. A webpage is parsed by a browser to obtain its Document Object Model (DOM) Tree structure. The DOM Tree is examined to discover different DOM Tree patterns for each class of Web image.

*Unlisted images* are standalone or random images that appear anywhere on a page (*c.f.* Fig 1a: Segment 9), for example, profile photos in personal homepages, company logos, advertisements etc. The corresponding DOM Tree for such images and their surrounding context is consistently an image node with its surrounding text as text node siblings, with a root HTML tag representing the boundary of this image segment (*c.f.* Fig 1b).

*Listed images* are two or more images that are systematically ordered within the webpage (*c.f.* Fig 1a: Segment 1-8). Examples of listed images are representative images, list of product images, news images, etc. The associated DOM Trees for such image segments are characteristically the image node with its surrounding text nodes that are a sub-tree under a root HTML tag defining the segment boundary. Other siblings under this root HTML tag share similar sub-tree structure (*c.f.* Fig 1d).

*Semi-listed images* are visually similar to listed images. The difference is characterized by their DOM tree. Their DOM tree is similar to a DOM Tree of an unlisted image in the sense the image node with its surrounding text nodes are under a root HTML tag that represents the segment boundary but along with those nodes, there are other image nodes with their own surrounding texts nodes as well on the same level (*c.f.* Fig 1c).

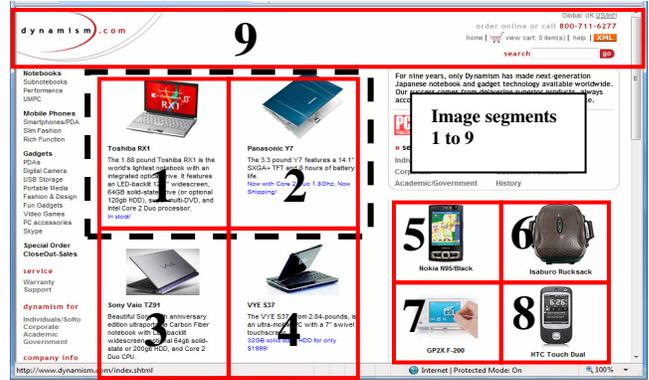
(a) Image segments 1 - 9

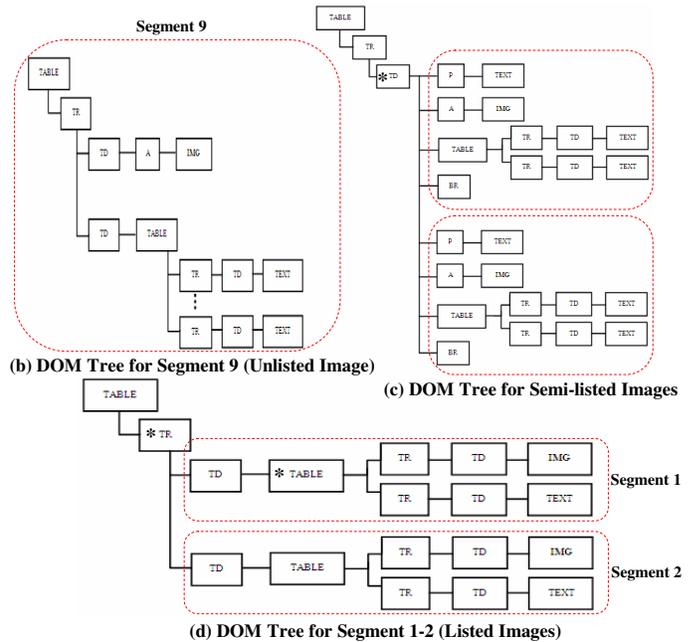
(b) DOM Tree for Segment 9 (Unlisted Image)
(c) DOM Tree for Semi-listed Images
(d) DOM Tree for Segment 1-2 (Listed Images)

**Figure 1. Example of image segments and their corresponding DOM Tree Structures**

Commonly, for all images, their surrounding context are texts in close proximity to the image within a webpage as well as in the webpage's DOM Tree structure, the corresponding text nodes are neighboring nodes to the image node in the DOM Tree, and all image nodes and text nodes are leave nodes in the DOM Tree.

### 3.2 Algorithm
We propose a novel DOM Tree based segmentation algorithm to extract image segments from webpages using the image characteristics mentioned above to determine the heuristics for segmentation. For every image node found in the DOM Tree, the algorithm finds the image segment using heuristic determined by the image characteristics. This is accomplished by detecting the variation in total number of texts in each upward level of the DOM Tree, beginning from the image node. We use Segment 1 from Fig. 1a to explain this, from the image node, the algorithm traverses up the DOM Tree, and stops at *<TABLE> node, when it first detects text nodes. An increase from zero text nodes to one text nodes as can be seen from the respective DOM Tree structure in Fig. 1d. The change denotes the first image segment, which is the solid rectangular box highlighting Segment 1. The algorithm continues upwards until it detects another change in the number of text nodes at *<TR> node, a bigger segment is detected as shown in the dashed box encompassing Segment 1 and 2 in Fig 1a. Sibling sub-trees are checked for listed and unlisted images, the smaller segment is regarded as the boundary of surrounding context for listed image and the larger segment for unlisted image. Sub-trees containing listed images will have siblings with similar sub-tree structure. In our example, the image is a listed image; therefore the smaller segment, which is the solid rectangular box, is taken as the segment boundary for segment 1.

```
Algorithm: The Segmentation Algorithm for Web Images
Require: I ← The set of valid image nodes from a webpage.
1:  for all i_k ∈ I do
2:    repeat
3:      int stateImg = getNumImage();
4:      int stateText = getNumTextNode();
5:      int state = 0;
6:      if (stateText!=state && stateImg>0 && stateText>0)
7:        if (stateChangeTwice)
8:          if (parentOfListedImage)
9:            take childNode as region
10:         else (parentOfUnlistedImage)
11:           take parentNode as region
12:         end if
13:       else
14:         if (SectionsInSameLevelForSemiListedImage)
15:           partition sections accordingly
16:         else
17:           state = stateText;
18:           stateChangeTwice = true;
19:         end if
20:       end if
21:       childNode = parentNode;
22:       parentNode = parentNode.getParentNode();
23:     end if
24:   until parentNode != null
25: end for
```

The algorithm is iterative. It starts with the image node to identify image segments and stops when all valid images are processed. An image segment must contain at least one image node and one text node. The *stateImg* and *stateText* variables keep track of the number of image and text nodes respectively, while *state* variable records the current state. Our algorithm detects segments, by comparing the *stateText* variable to the *state* variable. The initial segment that is smaller and its bigger super-segment are detected by the *stateChangeTwice* variable. The image characteristics influence the decision to take which segment as the finalized image segment.

Upon detection of the initial segment, our algorithm checks for semi-listed images whose sections occur in the same level of the DOM Tree (*c.f.* Fig. 1c). We search for patterns and separation point that divide the semi-listed images and their surrounding context. If these sections exist, we proceed to partition them into individual sections and extract them accordingly. For example, in Fig. 1c, the initial section is detected at node *<TD>, we check for repeating patterns in this sub-tree and here, the nodes repeat themselves in the sequence of <P>, <A>, <TABLE> and <BR>. Upon identifying the sequence, the separation points is determined to partition them according to regions; in this example, the starting cutoff point is set to node <P> and an ending cutoff point is set to node <BR>. Consequently, the semi-listed images are partitioned into two sections shown in the dashed boxes. Otherwise, the algorithm will resume upward traversal until another change in the number of text indicating that a bigger section is found. At this point (*i.e. stateChangeTwice* is true), we check for listed and unlisted images. If listed images are found, we take the smaller section as the image segment; otherwise, we take the bigger section as our choice.

Therefore, based on the variations found in the number of text nodes at different DOM Tree levels, our algorithm detects image segments and extracts them accordingly.

## 4. EXPERIMENTS AND RESULTS
In this section, we present an empirical evaluation of our segmentation algorithm for Web images.

### 4.1 Contribution of a Human-labeled Dataset
A user study is conducted to address the problem of ambiguity in defining the boundary of the contextual information for web image. 30 subjects were recruited to perform manual segmentation *i.e.* to identify all images and their surrounding context, on 100 randomly selected webpages across various categories in Alexa Web Directory such as news, business, shopping, health, entertainment, people and society. Each subject had 10 random webpages to segment; therefore, each webpage was at least segmented by 3 subjects, resulting in 3 sets of data. The recruited subjects were students and lecturers from local universities; 13 males and 17 females.

The outcome is a human-labeled dataset comprising of 1019 image segments. Fig 1a shows examples of the users' perception of image segments, labeled from 1-9. Indisputably, they regarded images and their associated textual information as sections within a webpage. When identified sections differ between subjects, we chose to consider the bigger section rather than the section that has been defined by at least two subjects. By accepting the bigger sections, we do not lose out on the general topic header relevant to the smaller sections.

From this study, the size of a valid image can be clearly defined. Most work discarded images with both width and height less than 60 pixel and width-height ratio less than 1/5 or greater than 5 [1, 6, 8]. In the study, users, on top of that, identified images with both height and width less than 60 pixel but greater than 45 pixels and provided that these images are square or rectangular in shape *i.e.* width-height ratio between ½ and 2. Hence, in addition to the valid image size defined in the literature, we will also extract image segments containing square/rectangular images with width and height of between 45 and 60 pixels.

The resulting image segments are consistent as verified using the "Split-half Analysis Consistency". The average Pearson's correlation value is computed for the 3 sets, which is equal to 0.93, indicating that the data obtained is highly consistent.

### 4.2 System-based Evaluation
We evaluate our segmentation algorithm within a system-based framework where the Precision and Recall indicators are used. Precision is the percentage of correctly extracted segments over the total extracted segments and recall is the percentage of correctly extracted segments over the total actual number of image segments in the dataset. We define *actual* as the images with the expected surrounding context; *extracted* as the images and their surrounding context extracted by the algorithm; and lastly, *correct* as the extracted images and their surrounding context that match the expected ones in *actual*.

The webpages are parsed using the HTMLParser available at http://htmlparser.sourceforge.net/ to obtain their DOM Tree. Our segmentation algorithm is then performed on the resulting DOM Trees, a total of 1012 image segments are *extracted*, slightly less than the *actual* 1019 segments. 748 segments are *correct*, thus, achieving **73% for both precision and recall**. The average time taken to process a webpage, extract the images and their contextual information, is **0.4s**, evaluated on a hardware platform with Duo Core 1.7GHz Intel Pentium Processer and 1GB RAM.

We compare our method to the Vision-based Page Segmentation (VIPS) algorithm, a heuristic DOM-based segmentation algorithm with additional visual information such as horizontal and vertical separators obtained from rendering the webpage. Each resulting visual segment has a defined Degree of Coherence to measure the content consistency within the block, ranging from 1 – 10. The greater the value, the more consistent the content is within the segment. An adjustable pre-defined Permitted Degree of Coherence (PDoC) value is provided to achieve different granularities of content structure to cater for different applications [2]. VIPS is selected as its executable is available at http://www.cs.uiuc.edu/homes/dengcai2/vips/vips.html and it is widely used in many web-based image retrieval systems to extract the image contextual information. For our comparison purposes, we emulate He's work in applying VIPS to extract images and their surrounding texts whereby the PDoC value is empirically set to 5 [8]. Several pages could not be segmented by VIPS due to scripting error and therefore they are excluded from the test. The result is tabulated below.

**Table 1: Performance Comparison using VIPS and our proposed method**

|           | Our Method | VIPS=5 | VIPS=6 | VIPS=7 |
|-----------|------------|--------|--------|--------|
| Actual    | 869        | 869    | 869    | 869    |
| Extracted | 864        | 853    | 853    | 853    |
| Correct   | 628        | 174    | 278    | 333    |
| **Recall**    | **0.72**   | **0.20** | **0.32** | **0.38** |
| **Precision** | **0.73**   | **0.20** | **0.33** | **0.39** |

The table clearly illustrates that our method outperformed VIPS in extracting image segment across a diverse assortment of webpages, mainly because the PDoC value of 5 is not the most optimal value for VIPS to extract image segments. It should be noted that the image segments are considered correct only if the right amount of image contextual information is extracted, no more and no less. VIPS performed poorly because at PDoC set to 5, VIPS tends to take the bigger section as an image segment, referring back to Fig 1a, segments 1-4 are considered as one image segment and segments 5-8 as another image segment. If the bigger segments are considered *correct*, then both the algorithms would have achieved over 90% precision. However, bigger sections usually contain lots of noise that is meaningless to the image. Hence, we further test VIPS for higher PDoC value, as shown above. Result stops at PDoC=7 as many pages are too finely segmented until the surrounding context for the images are less than in the *expected* image segments. We find the PDoC value poses a problem for image segmentation especially for webpages with multiple arrangements of images or across a range of diverse webpages.

This initial report of performance studies verified the effectiveness of our segmentation algorithm. We believe the precision of the algorithm can be further increased if it were able to tackle deeply structured DOM Trees as in blogs, which have a list of posts containing images. Such images are listed images, however, the repetitive sub-tree patterns are found beyond the two changes in the number of texts.

## 5. DISCUSSION & CONCLUSION

This paper introduced a fully automated segmentation algorithm without any tuning parameters. It is a DOM Tree-based segmentation algorithm without going through underlying browser rendering engine to obtain visual cues. Even without visual cues, it is able to extract an image and its contextual information efficiently. Segments that do not contain any image are discarded. For every image extracted, the segmentation method only searches through the surrounding region, thus making it more efficient and scalable for large web sites containing huge amount of images. Our dataset might be small but a variety of webpages was included and it is manually established by a group of users, our future work is to test the algorithm more extensively as well as to extend it to cater to webpages with deep DOM Tree structures.